\documentclass[superscriptaddress,twocolumn,showpacs,preprintnumbers]{revtex4}
\usepackage{amsmath}
\usepackage{graphicx}
\usepackage{dcolumn}
\usepackage{bm}
\usepackage{subfigure}

\begin{document}

\title{The Dynamics of Sustained Reentry in a Loop Model with Discrete Gap
Junction Resistance}
\author{Wei Chen}
\email{wei.chen@umontreal.ca}
\affiliation{Department of Physioloy, Institute of Biomedical Engineering, Universit\'{e} 
\\
}
\author{Mark Potse}
\email{mark@potse.nl}
\affiliation{Department of Physiology, Institute of Biomedical Engineering, Universit\'{e}
de Montr\'{e}al, Canada and Centre de Recherche de l'H\^{o}pital du Sacr\'{e}%
-Coeur de Montr\'{e}al }
\author{Alain Vinet}
\email{alain.vinet@umontreal.ca}
\affiliation{Department of Physiology, Institute of Biomedical Engineering, Universit\'{e}
de Montr\'{e}al, Canada and Centre de Recherche de l'H\^{o}pital du Sacr\'{e}%
-Coeur de Montr\'{e}al\\
}
\date{\today }

\begin{abstract}
Dynamics of reentry are studied in a one dimensional loop of model cardiac
cells with discrete intercellular gap junction resistance ($R$). Each cell is
represented by a continuous cable with ionic current given by a modified
Beeler-Reuter formulation. For $R$ below a limiting value, propagation is
found to change from period-1 to quasi-periodic ($QP$) at a critical loop length ($L_{crit}$) that decreases with $R$. Quasi-periodic reentry exists from $L_{crit}$ to a minimum length ($L_{min}$) that is also shortening with $R$. The decrease of $L_{crit}(R)$ is not a simple
scaling, but the bifurcation can still be predicted from the slope of the
restitution curve giving the duration of the action potential as a function
of the diastolic interval. However, the shape of the restitution curve
changes with $R$.
\end{abstract}

\maketitle





\section{Introduction}

Self-sustained propagation of electrical activity around a one-dimensional
(1-D) loop of cardiac tissue is the simplest model of reentry, the mechanism
by which a propagating activation front maintains itself by travelling
around a functional or anatomical obstacle. Reentry has been much studied
because it was demonstrated to be an important mechanism of cardiac
arrhythmia.\cite{bl, bj, fl, rp} For the 1-D loop, most work has been done
assuming the membrane to be a continuous and uniform cable with constant
intracellular axial resistivity.\cite{va1, va2, va3,va4, karma94, cm,
mc2,ka, lt, qw,gray2005} For different models representing the ionic
properties of the membrane, propagation was found to change from stable
period-1 propagation to quasi-periodic reentry when the length of the loop
was reduced below a critical length. The quasi-periodic reentry was
characterized by a spatial oscillation of the action potential duration as
propagation proceeded around the loop. Based on numerical simulation, the
bifurcation was in most cases classified as supercritical, with the
amplitude of the oscillation growing as the length of the loop was reduced
below the critical length. Quasi-periodic reentry was found to exist from
the critical length to a minimal length below which sustained propagation
became impossbile. In some instances, two different modes of quasi-periodic
propagations were identified, with different wavelengths, different interval
of existence, and sometimes different scenarios of creation.\cite{va2,
va3,va4, cm,mc2} Different attempts were made to build simplified
representations of the dynamics allowing analytical examination of the
nature of the bifurcation.\cite{mc2,va3,cp2,karma94,keener_2002,gootwald_2006} One of these approaches,
which guides the present investigation, relies on an integral-delay model.
\cite{cm, mc2} It is based on the assumption that both the speed of
propagation and the action potential duration can be expressed as functions
of the diastolic interval, which measures the recovery time from the end of
the previous action potential. The model has been successful in reproducing
the locus of the bifurcation observed by numerical simulations of 1-D loops
with Beeler-Reuter type representation of the membrane. It predicts that the
bifurcation should occur when the diastolic interval in the period-1 reentry
reaches the critical value where the slope of the restitution curve becomes
1.

However, cardiac excitable tissue is not a syncytium, but rather a mesh of
myocytes connected by discrete gap junction resistance.\cite{spach2000} Much
work has been done to investigate the effect of discrete gap junction
resistances in a one-dimensional structure,\cite{keener1991,ry1,qw, vf,
yr,rs} mostly focused on the effect of resistivity on excitability. In the
discrete case, the resistance no longer acts as a scaling factor with regard
to space. Because the intercellular current is reduced as the gap junction
resistance is increased, the latency of the cell to cell propagation is
increased until propagation fails at some limiting value of the resistance.
Besides, upon premature or repetitive stimulations, the excitability of the
tissue must be more recovered for propagation to proceed, which corresponds
to an increase of the refractory period.

This paper describes how the bifurcation from period-1 to quasi-periodic
propagation and the characteristics of the quasi-periodic propagation are
modified by the increase of the intercellular resistance in a 1-D loop of
discrete model cardiac cells. This paper is organized as follows. In the
next section, the model and computational method are described. The results
of the numerical simulation are presented in \S 3. The bifurcation from
stable period-1 reentry is explained in \S 4. The $QP$ modes of reentry are
analyzed in \S 5. The final section is devoted to a summary and discussion.

\section{Methods}

We consider a one-dimensional loop formed by $N$ identical cells connected
by gap junction resistanced. Each cell is modeled as a continuous and
uniform cable of radius ($a$) 5 $\mu m$, length ($L_{c}$) 100 $\mu m$ and
intracellular resistivity ($\rho $) 0.2 $K\Omega $$\cdot $$cm$ lying in an
unbounded volume conductor of negligible resistivity. The transmembrane
potential ($V^{i=1,N}$ in mV) of the cells is described by the well-known
cable equation:

\begin{eqnarray}
\label{model}
\frac{1}{\rho }\frac{\partial ^{2}V^{i}(x,t)}{\partial x^{2}} &=&S[C_{m}
\frac{\partial V^{i}(x,t)}{\partial t}+I_{ion}^{i}(x,t)],  \nonumber \\
x &\epsilon &\left\{ 0,L_{c}\right\} ,\hspace{0.1cm}i\hspace{0.1cm}\epsilon 
\hspace{0.1cm}\left\{ 1,N\right\}
\end{eqnarray}
in which $C_{m}$ is the membrane capacitance (1 ${\mu }F/cm^{2}$), $S$ is
the surface-to-volume ratio (0.4 ${\mu }m^{-1}$) and $I_{ion}$ is the ionic
current ( ${\mu }F/cm^{2}$). The membrane ionic model is the same MBR model
that was used in our previous works on continuous 1-D and 2-D rings.\cite
{va2,va3,va4,rf,cp} In this model, the sodium current is controlled by two
inactivation gate variables $h$ and $j$. Each cell is connected to its
neighbors by a discrete gap junction resistance $R$ ($K\Omega $). Continuity
of the intracellular current between the cells yields the boundary
conditions: 

\begin{eqnarray}
\label{boundary}
\frac{\partial V^{i}}{\partial x}|_{x=L_{c}} &=&\frac{\partial V^{mod(i,N)+1}
}{\partial x}|_{x=0}=-\frac{\rho }{{\pi }a^{2}}I_{i,mod(i,N)+1} \nonumber\\
V^{i}(L_{c}) &-&V^{mod(i,N)+1}(0)=RI_{i,mod(i,N)+1}
\end{eqnarray}

For simulation, we have modified the numerical method that was developed for
continuous loops.\cite{va2} Briefly, for each time step (${\Delta }t=$ $2{\
\mu }s)$, Eq.(\ref{model}) becomes equivalent to an ordinary differential
equation

\begin{equation}
\frac{d^{2}V^{i}(x)}{dx^{2}}-K^{2}V^{i}\left( x\right) =g^{i}(x)
\label{model2}
\end{equation}
whose solution can be expressed as the sum of a particular solution $V_{p}^{i}\left( x\right)$ and of the homogeneous solution

\begin{equation}
V_{h}^{i}(x)=A_{i}e^{kx}+B_{i}e^{-kx}  \label{model2_vh}
\end{equation}

$V_{p}^{i}\left( x\right) $ is obtained by solving Eq.(\ref{model2}) with
Neumann boundary conditions (${\partial }V^{i}/{\partial }x|_{x=0,L_{c}}=0$)
using the linear finite element method\cite{lt} with a uniform grid (${\
\Delta }x=25{\mu }m)$, i.e. 5 nodes. Cells are then reconnected by choosing
the coefficient of the homogeneous solutions to fulfill the continuity
conditions Eq.(\ref{boundary}). For a subset of $R$ values, calculations
repeated with ${\Delta }x=12.5{\mu }m$ and ${\Delta }t=1{\mu }s$ gave the
same results.

The purpose of the simulations is to obtain a description of the regimes of
reentry of the function $R$ and $L=NL_{c}$, the length of the loop. During
reentry, the successive action potentials ($j=1,m$) at each node can be
characterized by their activation times ($T_{act}^{j}$), set at the maximum
derivative of the upstroke, and their repolarization times ($T_{repol}^{j}$
), taken at the $-50$ mV downcrossing in repolatization. The action
potential duration ($A$) and the diastolic interval ($D$) associated to each
action potential are calculated respectively as $A^{j}$ = $
T_{repol}^{j}-T_{act}^{j}$ and $D^{j}$ = $T_{act}^{j}-T_{repol}^{j-1}$.\cite
{va1,va2,va3,va4} The propagation of the wavefront along the loop generates
spatial profiles of $A$ and $D$ that typify the reentry. In contrast to a
continuous loop, propagation on a discrete loop can be patterned inside each
cell but identical across all the cells. We have chosen to use only the $A$
and $D$ values of the middle node of all cells to characterize the reentry.
We label period-1 (P-1) reentries in which $A$ and $D$ remain constant
across all the middle nodes, and quasi-periodic reentries where $A$
and $D$ oscillate both in time and space.

For each value of $R$, an initial $L$ was chosen large enough to sustain P-1
stable reentry. Reentry was initiated by transiently opening the loop and
stimulating one end. Computation was continued until stable period-1 reentry
was detected, the stability criteria being less than 0.5 ms difference in $A$
and $D$ between all the middle nodes for one rotation of the front.
Afterward, the loop length was gradually reduced by steps of one cell, using
the final state of the previous $L$ as initial condition and removing one
cell far from the position of the excitation front. When the stability
criterion was not fulfilled after a minimum of 25 turns, reentry was
labelled as $QP$. With this procedure, both $L_{crit}$ and $P_{
crit}$, respectively, the minimum length and minimum period with P-1
reentry, as well as $L_{min}$ , the minimum length for sustained
reentry, were identified for each value of $R$. In some instances,
bistability between P-1 and $QP$ reentry was investigated by stepwise
expanding loops that were initially in quasi-periodic regime. One cell was
inserted in the loop, with initial conditions set at the mean of the states
of its neighboring cells. Finally, we also searched for distinct modes of $
QP $ reentry using the method described in \cite{va3}, in which the $D$
spatial profile of a $QP$ solution for a given $L$ value is compressed by a
scaling factor and used to construct an initial condition to find
alternative $QP$ solutions with smaller wavelengths.

\section{Results}

\begin{figure}[<h>]
\centerline{\includegraphics[width=70mm]{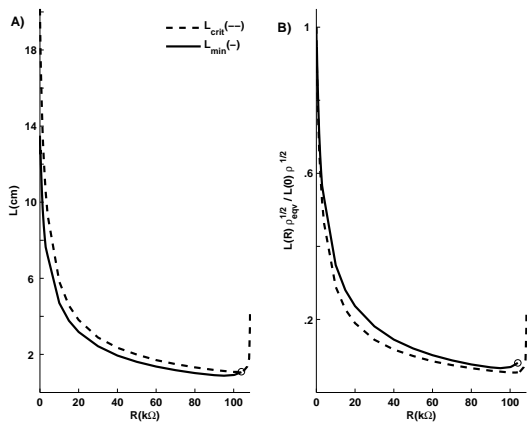}}
\caption{A) $L_{crit}$ (cm, dashed line), the shortest $L$ with period-1 reentry, and $L_{min}$ (cm, continuous line), the mininum $L$ with $QP$ reentry, as a function of the gap resistance $R$($K{\Omega}$). B) Normalized values of $L_{crit}$ and $L_{min}$ (see text) as a function of $R$.}
\label{Fig1}
\end{figure}

Figure \ref{Fig1} A) shows $L_{crit}$ and $L_{min}$ as a
function of $R$. Both $L_{crit}$ and $L_{min}$ decrease until
they merge at $R=$ 104 $K{\Omega }$. From this resistance, $QP$ reentry does
not exist anymore and P-1 reentry remains the only regime of sustained
propagation. From there, the limiting length for P-1 reentry \ increases
until sustained propagation becomes impossible at $R=108.429K{\Omega }$.
Increasing the resistivity in a continuous loop would also decrease $L_{
crit}$ and $L_{min}$. However, the speed of propagation being
proportional to $1/\sqrt{\rho }$ in a continuous media,\cite{kleber} $\sqrt{
\rho }L_{crit}(\rho )$ and $\sqrt{\rho }L_{\min }(\rho )$ would
remain invariant. To compare the continuous and discrete medium, we computed
the equivalent resistivity of the latter as

\begin{equation}
\rho _{eqv}(R)=\rho +\frac{NR\pi a^{2}}{L}=\rho +\frac{R\pi a^{2}}{
L_{c}}  \label{equv}
\end{equation}

If the two media were equivalent, the ratio $L(R)\sqrt{\rho _{eqv}(R)}
/L(0)\sqrt{\rho }$ would remain equal to 1. Fig. \ref{Fig1} B) shows clearly
that the diminution of $L_{crit}$ and $L_{min}$ cannot be
explained by a simple scaling, as it occurs in a continuous medium.

\section{$L_{crit}$ , $P_{crit}$ in transition to $QP$ reentry.}

\begin{figure}[<h>]
\centerline{\includegraphics[width=70mm]{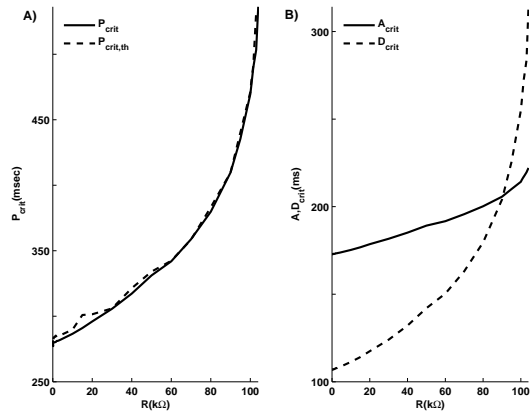}}
\caption{A) Continuous line: For each value of the gap resistance $R$, the critical cycle length $P_{crit}$ ($ms$) at $L_{crit}$, the shortest loop with period-1 reentry. Dashed Line: $P_{crit,th}$, the critical cycle length computed from the restitution curve (see text). B) Value of the action potential duration ($A_{crit}$, continuous line) and diastolic interval ($D_{crit}$, dashed line) at $L_{crit}$.}
\label{Fig2}
\end{figure}

\begin{figure}[<h>]
\centerline{\includegraphics[width=70mm]{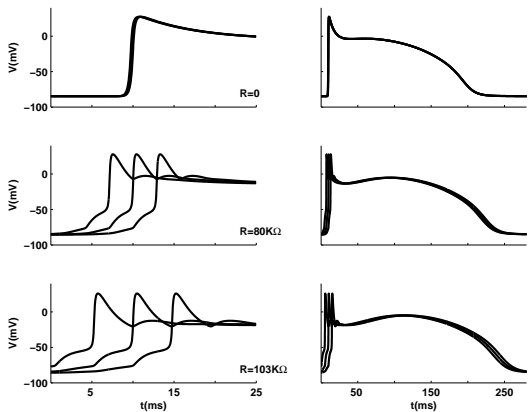}}
\caption{Action potentials ($mV$) in the first node of 3 successive cells as a function of time ($ms$) during period-1 reentry for, from top to bottom, $R=$ 0,800 and 103 $K{\Omega}$. On the left column panels, only the activation is shown, while the complete action potentials are displayed in the rigth column panels.}
\label{Fig3}
\end{figure}

Two distinct scenarios can lead to the disappearance of P-1 reentry. For $
108.429K{\Omega }>R>104K\Omega $, sustained reentry does not exist for $
L<L_{crit}=L_{min}$, so that reentry ends abruptly with the disappearance of
the P-1 solution. For $R<104K\Omega $, P-1 reentry is replaced by $QP$
reentry that persists from $L_{crit}$ to $L_{min}$. In this
section, we consider the second type of transition. In the continuous MBR
loop, the bifurcation from P-1 to $QP$ propagation was shown to occur at the
critical period $P_{crit}=D_{crit}+A_{crit}$ where $D_{crit}$ and $A_{crit}$ are the values for which the slope of
the restitution curve $A(D)$ reaches 1.\cite{va2, cm} $P_{crit}$ is
constant and independent of $\rho $ in a continuous medium. In contrast, Fig.
\ref{Fig2} A) shows that $P_{crit}$ increases with $R$ in the
discrete loop. Both $A_{crit}$ and $D_{crit}$ contribute to
the change of $P_{crit}$ (Fig.\ref{Fig2} B), but the increase of $D_{
crit}$ is more important. For each value of $R$, we collected the $D$
and $A$ values of the P-1 solutions for a collection of $L$ values as well
as those of the first $QP$ solution below $L_{crit}$ to construct the 
$A(D)$ restitution curve. Each curve was fitted with a simple exponential to
find $D_{crit,th}(R)$, the value where the slope of the fitted $
A(D)=1 $, and theoretical value $P_{crit,th}=D_{crit,th}+A(D_{crit,th})$. As shown in Fig.\ref{Fig2} A), $P_{crit,th}$ falls
very close to the $P_{crit}$ values found by simulation. Hence, the
mechanism responsible for the transition from P-1 to $QP$ reentry is the
same in the continuous and discrete loop, and the increase of $P_{crit}$ results from $R$ transforming the restitution curve. The mechanisms
responsible for the change of $D$ and $A$ can be identified in Fig. \ref
{Fig3}, which shows the action potentials of the first node in the three
successive cells for increasing values of $R$. (top to bottom, $R=0$, 80 and
103 K$\Omega $). Increasing $R$ prolongs the latency of the action potential
(left column panels). Because the end of the diastolic interval is set at
the maximum upstroke derivative, the increase of latency is translated as an
increase of $D$. On the other hand, neighboring cells exchange current
during the repolarization, which compensates for the delay of activation and
prolongs the action potentials. These two effects contribute together to the
change of the restitution curve. Finally, the transition from period-1 to $
QP $ solution was always found to be supercritical, as in the continuous
case.

\begin{figure}
\centerline{\includegraphics[width=60mm]{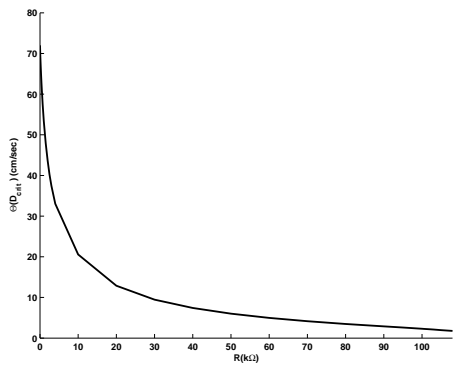}}
\caption{Intercellular activation speed (calculated between the first node of successive cells, $cm/sec$) at $L_{crit}$ as a funtion of the gap resistance $R$.}
\label{Fig4}
\end{figure}

Once $P_{crit}$ is known, $L_{crit}$ can be calculated if the speed
of propagation $\Theta (D_{crit})$ is provided. In the discrete
media, the total time to propagate from one cell to another is a composite
of propagation time within and between the cells. The former decreases with $
R$ , while the latter, which is equivalent to the latency displayed in Fig. 
\ref{Fig3}, increases. The final composite $\Theta (D_{crit})$ is
shown in Fig. \ref{Fig4}.

\section{$QP$ reentry}

The characteristics of the $QP$ reentry in the continuous MBR loop have been
extensively discussed in previous papers.\cite{va2} Two modes of $QP$ were
identified, characterized by $D$ oscillations with different spatial
wavelengths ($\lambda $). The first mode, referred to as mode-0, exists from 
$L_{crit}$ to $L_{min}$. Its $\lambda $, close to two turns of
the loop at $L_{crit}$, diminishes as the loop is shortened, but
always remains longer than $L$. The second, referred as mode-1, exist only
over a subset of the $\left[ L_{min},L_{crit}\right] $
interval with a $\lambda $ always less than $L$. These two types of $QP$
solutions were found for all values of $R<104K\Omega $ where $QP$ solutions
exist.

\subsection{mode-0 $QP$ reentry}

\begin{figure}[<h>]
\centerline{\includegraphics[width=70mm]{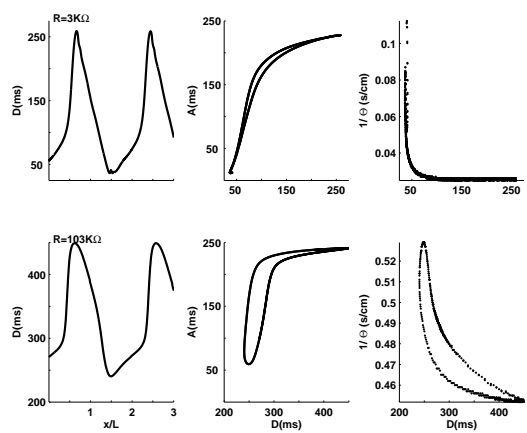}}
\caption{Characteristics of the mode-0 $QP$ solution at $L_{min}$ for $R =$ 3$K\Omega$ (top row panels) and $R =$ 103$K\Omega$ (bottom row panels). Left panels: $D$, the diastolic interval) as a function of position ($x/L$) for 3 successive turns abutted end-to-end. Middle panels: $A$, the duration of the action potential, as a function of $D$, from the solutions shown in the left panels. Rigth Panels: $1/\Theta$, the cell-to-cell conduction time, as a function of $D$ from the solutions shown in the left panels. Only the data of the first node of each cell were used to construct these plots.}
\label{Fig5}
\end{figure}

\begin{figure}[<h>]
\centerline{\includegraphics[width=70mm]{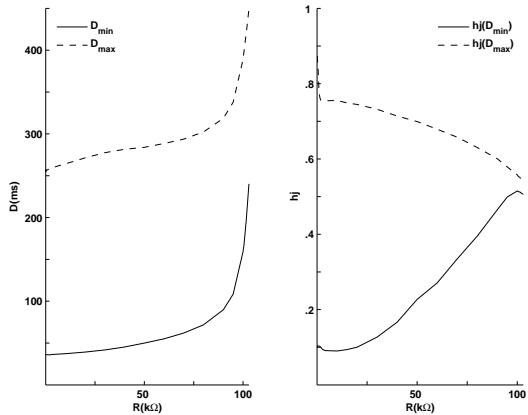}}
\caption{Left Panel: $D_{min}$ (continuous line) and $D_{max}$ (dashed line) respectively the minimum and maximum diastlolic interval of the mode-0 $QP$ solutions at $L_{min}$ as a function of $R$. Rigth Panel: $hj$, product of the sodium current inactivation gates taken at $D_{min}$ (continuous line) and $D_{max}$ (dashed line) for the mode-0 $QP$ solutions at $L_{min}$.}
\label{Fig6n}
\end{figure}

We first consider the mode-0 solutions that exist over the whole $\left[ L_{
min},L_{crit}\right] $ interval. Fig. \ref{Fig5} presents the
characteristics of the mode-0 solutions at $L_{min}$ for two values
of $R$ (top panels , $R=3K\Omega $, $L=7.65cm$ , bottom $R=103K\Omega $ , $
L=1.04cm$ ). The leftmost panels show the spatial oscillation of $D$ by
plotting successive turns end to end. The first obvious difference is the
range of $D$ values covered by the two solutions. The left panel of Fig. \ref
{Fig6n} shows $D_{min}$ and $D_{max}$, the minimum and maximum
value of $D$ for the mode-0 solution taken $L_{min}$ as a function of 
$R$. It is well known that, in a discrete medium, the minimum excitability
needed to sustain propagation increases as a function of $R$ until a
limiting value beyond which propagation is blocked even in a medium at rest.\cite{vf} In the MBR model, the excitability can be measured by the product $
hj$ of the inactivation gates of the sodium current. The right panel of Fig. 
\ref{Fig6n} shows $hj(D_{min})$ and $hj(D_{max})$, the
excitability of the action potentials produced respectively at $D_{min}$ and $D_{max}$ for the mode-0 solutions at $L_{min}$. As $R$
increases, the minimal excitability allowing propagation becomes higher,
which requires an increase of $D_{min}\left( R\right) $. At $
R=104K\Omega $, $hj(D_{min})=hj(D_{crit})$, $QP$ propagation
disappears and only P-1 reentry remains. On the other hand, the curve $
hj(D_{max})$ rather reflects the inactivation of the socium current
occurring during the latency preceding the upstroke of the longer action
potential. At $R=104K\Omega $, the limit for $QP$ propagation, $hj(D_{
max})>hj(D_{min})=hj(D_{crit})$, which indicates that P-1
propagation is still possible if $R$ is increased. However, the difference
is small, such that the range of $R$ values over which P-1 reentry can still
occur is limited, as it is seen in Fig.\ref{Fig1}.

The middle column panels of in Fig.\ref{Fig5} display the $A(D)$ relation
obtained from each $QP$ mode-0 solution. Each curve has two branches, the
lower and upper branch coming respectively from increasing and decreasing
portion of the $D$ spatial profile. Such a dual structure has been observed
in the continuous loop and was explained either by the influence of
neighbors on the repolarization \cite{cp2} or by short term memory \cite
{gray2005}. The separation between the branches is enhanced by the increase
of $R$. Finally, the right column panels of Fig.\ref{Fig5} show $1/\Theta $
vs $D$, the dispersion relation of the conduction time. For $R=3K\Omega $
(top right panel), the dispersion relation appears as a single value
function, similar to what is seen in the MBR continuous loop. For $
R=103K\Omega $ (bottom right panel), the dispersion relation has two
branches, as the $A(D)$ curve. The lower branch is associated with the
decreasing portion of the $D$ spatial profile.

\subsection{Higher $QP$ Modes}

\begin{figure}[<h>]
\centerline{\includegraphics[width=70mm]{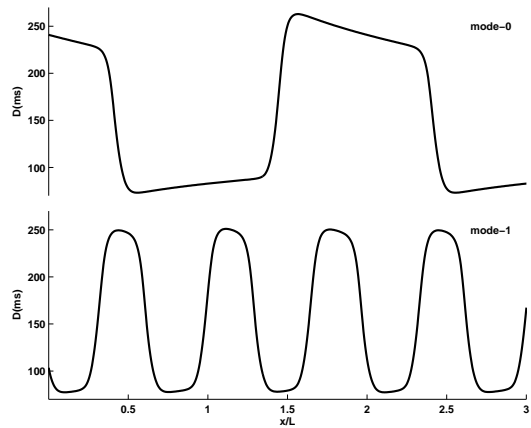}}
\caption{Mode-0 (top panel) and mode-1 (bottom panel) $QP$ solutions for $L=$1.80$cm$ and $R=$50$K\Omega$. The plots show $D$, the diasolic interval, as a function of position ($x/L$) for three turns abutted end-to-end.}
\label{Fig7}
\end{figure}

Fig.\ref{Fig7} \ shows an example of mode-0 and mode-1 solution for $
R=50K\Omega $ and $L=1.8$ cm, in the middle of the $\left[ L_{min},L_{
crit}\right] $ =$\left[ 1.61cm,1.99cm\right] $ interval for this
value of $R$. Mode-1 solutions were found for all value of $R$ with $QP$
propagation over a subset of the $\left[ L_{min},L_{crit}
\right] $ interval, as in the case of the continuous cable. Courtemanche et
al. \cite{mc2} in their analyses of a delay-integral model representing
reentry on a 1-D loop have predicted the existence of an infinite number of $
QP$ modes, with spatial wavelengths near $L_{crit}$ given by

\begin{equation}
\lambda (n)=\frac{2L}{2n+1}-\frac{C}{(2n+1)^{3}}
\end{equation}
where $n$ is the order of the mode and C is a small positive constant. As
seen Fig.\ref{Fig7}, $\lambda (0)/\lambda (1)$ is indeed close to 3.
However, in the MBR continuous loop, only the first 2 modes (i.e. 0 and 1)
were observed. This was explained by the effect of resistive coupling
between neighbors that limits the spatial gradient of voltage and forbids
the appearance of higher modes.\cite{va3} Theoretically, 
\begin{equation}
\frac{\lambda (2)}{\lambda (0)}\simeq \frac{1}{5},\hspace{1cm} \frac{
\lambda (2)}{\lambda (1)}\simeq \frac{3}{5}.
\end{equation}

To look for mode-2 solutions for different $R$ and $L$ values, we have
compressed the $D$ profiles of the mode-0 and mode-1 solutions up
respectively to a factor 6 and 2 to build different initial conditions. This
procedure was successful in obtaining mode-1 solutions from mode-0
solutions, but higher modes of propagation were never produced. For all
scaling factors, propagation was found to stabilize either to mode-0 or
mode-1.

\bigskip

\section{Discussion and Summary}

Increasing $R$ in the discrete loop allows sustained reentry to be
maintained in much shorter circuits than in continuous loops with equivalent
lumped resistance. The locus of the bifurcation from period-1 to $QP$
propagation can still be predicted from the $A(D)$ dispersion curve
constructed by gathering data from P-1 solutions and from mode-0 $QP$
solutions close to the supercritical bifurcation. However, increasing $R$
modifies $A(D)$ and the value of $P_{crit}$. On one hand, the latency
of the cell to cell propagation is increased, due to the decrease of the
intercellular current. This prolongs $D$ because it includes the latency and
shifts $A(D)$ to the left. On the other hand, once all neighboring cells are
activated, they exchange current that decreases the differences in potential
induced by the latencies. That prolongs $A$ and, as a consequence, lifts the 
$A(D)$ curve. Both effects act together to increase $P_{crit}=D_{crit}+A(D_{crit})$. In order to analyze these effects, it
would be more appropriate to separate the latency from the diastolic
interval, redefining $D$ from the end of the action potential to the minimum
of $V$ in repolarization and considering the latency $Lat$ from the end of $
D $ to the maximum derivative of the upstroke. Then both $A$ and $Lat$ could
be analyzed as functions of $D$ and $R$. However, even with this change, it
will be difficult to build a low dimensional equivalent model of the
propagation, extending the integral-delay model developed for the continuous
loop. As seen in Fig \ref{Fig5}, increasing $R$ enhances the dual structure
of $A(D)$ during propagation, which results from the effect of coupling in
repolarization. Moreover, the same type of dual structure also appears for $
1/\Theta $, which is almost equivalent to the latency at high $R$ values. It
thus becomes impossible to neglect the modulating effect of coupling on both 
$A$ and $\Theta $ at high $R$ values. Whether alternative approaches that
have been proposed for the continuous loop would be more appropriate remains
to be determined.\cite{karma94,keener_2002, gootwald_2006} In any case, we
are still far from a general low-dimensional model that could be applied in
situations including a dynamic change of the intercellular coupling, as in 
\cite{jr, ap,noma2006}.

$R$ also influences $L_{min}$, the minimal length with $QP$
propagation, because higher $R$ necessitates more excitability for
propagation. As a consequence, the minimum $D$ allowing sustained $QP$
reentry increases with $R$ until it reaches $D_{crit}(R)$. From this
value of $R$, $QP$ propagation becomes impossible. For $R$ above this
limiting value, period-1 reentry ends abruptly when its $D$ reaches the
minimal value allowing propagation. The minimal $L$ for propagation
increases until reaching the value of $R$ where propagation becomes
impossible even in a medium at rest. Again, the dynamics would be very
interesting to study in a medium with dynamic modulation of the gap
resistance.

In all cases with $QP$ propagation, we found the bifurcation from period-1
to mode-0 propagation to be supercritical. For some $R$ values, the nature
of the bifurcation was further ascertained by prolonging simulation up to
100 rotations. As in the continuous case, the mode-1 solutions were found to
exist in a subset on the $\left[ L_{min},L_{crit}\right] $
interval. We also devoted much effort to find $n>1$ modes of $QP$
propagation for different values of $R$, building initial conditions either
from mode-0 \ or mode-1 solutions for different $L$ within the $\left[
L_{\min },L_{crit}\right] $ interval. All these attempts were unsuccessful.
Our initial guess was that the increase of $R$ should allow more abrupt
gradients of potential to exist between the cells, thus permitting the
existence of higher modes of propagation. However, as seen in Fig \ref{Fig5}
, the dual structure of the $A(D)$ and $1/\Theta $ relations becomes more
pronounced at high $R$. This suggests that coupling still limits the
gradient below what would be needed for higher modes of propagation.

This work was supported by a grant of the Natural Sciences and Engineering
Research Council of Canada.

\end{document}